# Experimental Verification of Ill-defined Topologies and Energy Sinks in Electromagnetic Continua

David E. Fernandes[1], Ricardo A. M. Pereira[2], Sylvain Lannebère[1], Tiago A. Morgado[1],

Mário G. Silveirinha[3*]

[1]Instituto de Telecomunicações and Department of Electrical Engineering, University of Coimbra, 3030-290 Coimbra, Portugal

[2] Instituto de Telecomunicações and Department of Electronics, Telecommunications and Informatics, University of Aveiro, 3810-193 Aveiro, Portugal

[3]University of Lisbon, Instituto Superior Técnico, Avenida Rovisco Pais, 1, 1049-001 Lisboa, Portugal

E-mail: dfernandes@co.it.pt, r.pereira@ua.pt, sylvain.lannebere@co.it.pt, tiago.morgado@co.it.pt, mario.silveirinha@co.it.pt

**Abstract**

In this article, it is experimentally verified that nonreciprocal photonic systems with a continuous translation symmetry may have an ill-defined topology. The topological classification of such systems is only feasible when the material response is regularized with a spatial-frequency cutoff. Here, we experimentally demonstrate that inserting a small air gap in between two materials may effectively imitate an idealized spatial cutoff that suppresses the nonreciprocal response for short wavelengths and regularizes the topology. Furthermore, it is experimentally verified that nonreciprocal systems with an ill-defined topology may be used to abruptly halt the energy flow in a unidirectional waveguide due to the violation of the bulk-edge correspondence. In particular, we report the formation of an energy sink that absorbs the incoming electromagnetic waves with a large field enhancement at the singularity.

To whom correspondence should be addressed: E-mail: mario.silveirinha@co.it.pt

# I. Introduction

Since the discovery of the topological states of light by Haldane and Raghu over a decade ago [1, 2], topological photonics has been in the spotlight of scientific research. This flurry of attention is motivated by the extraordinary properties of topological materials: they do not support light states in the bulk region, but when terminated with an opaque boundary, they may support unidirectional edge states. The edge states are immune to backscattering and their propagation is impervious to small defects and imperfections of the interface. The topologically protected transport of light paved the way for novel and exciting applications in photonics [1-21].

The net number of edge modes in a topological system can be determined using the bulk-edge correspondence principle [3, 22-24]. Typically, the topological classification of photonic materials is done through the calculation of the Chern invariants, which are determined from the photonic band structure of the material and from the Bloch eigenstates [1, 2]. The calculation of the invariants usually requires the system to be periodic, so that the wave vector space is a closed surface with no boundary [25]. This is not the case of systems with a continuous translational symmetry, e.g. bulk materials modeled as continuous media, for which the wave vector lies in the unbounded Euclidean plane. Notably, some time ago it was shown that arbitrary bianisotropic materials or waveguides with a continuous translational symmetry can be topologically classified, provided the material response has a wave vector cutoff [25]. The cutoff may already be embedded in the physical response of the materials due to the nonlocality arising from charge diffusion [26-28], or other effects. Alternatively, a synthetic spatial cutoff can be enforced by inserting a small air gap in between the relevant materials [23, 25]. Without a suitable wave vector cutoff, the topology of a system with continuous translational symmetry is ill-defined and thereby the bulk-edge correspondence is inapplicable [23, 29].

It was theoretically shown in [21] that the violation of the bulk-edge correspondence creates the conditions to halt abruptly the edge mode propagation at a singular point. Thereby, ill-defined topologies create unique opportunities and may lead to unique physical phenomena, such as the concentration of an incoming wave at a single point with a massive field enhancement [21, 30-37]. Electromagnetic energy sinks are expected to be useful to enhance nonlinear effects [38-44] and for energy harvesting [45, 46].

Here, we investigate the propagation of topologically protected edge modes at the interface between a magnetized ferrite and a metallic wall. We numerically and experimentally verify that separating the ferrite and the metallic wall with a small air gap effectively imitates a spatial cutoff, and thereby regularizes the ferrite topology. In particular, for the waveguide with the air gap it is possible to correctly predict the emergence of a unidirectional edge state using the bulk-edge correspondence. On the other hand, without the air gap, the ferrite topology is ill-defined, and it turns out that when it is paired with the metal wall the system does not support edge states. Taking advantage of this peculiar property, we experimentally demonstrate the formation of an electromagnetic energy sink at the junction of materials with well-defined and ill-defined topologies. The time dependence of the electromagnetic fields is assumed of the type $e^{-i\omega t}$, with $\omega$ being the oscillation frequency.

## II. Topological classification and the bulk-edge correspondence

Here, we study the edge modes supported by a waveguide formed by a magnetically biased ferrite and a metal. The two materials are separated by a small air gap with thickness *d,* as shown in Fig. 1a. For now, it is assumed that the structure is uniform along the *z*-direction.

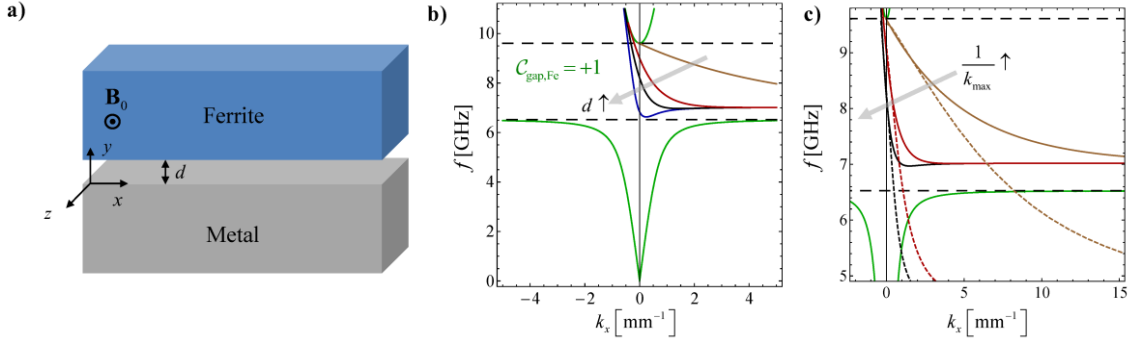

Fig. 1 **a)** Geometry of the edge waveguide consisting of a magnetically biased ferrite half-space and a metal half-space separated by an air gap with thickness *d*. **b)** Band diagram (TE waves) and gap Chern number of a lossless biased ferrite material with $\omega_l/(2\pi) = 4.43$ GHz and $\omega_g/(2\pi) = 5.18$ GHz (green solid lines). The bandgap is delimited by the horizontal dashed black lines. The figure also depicts the dispersion of the edge modes for the air gap thicknesses: $d = 0.1$ mm (brown solid curve); $d = 0.75$ mm (red solid curve); $d = 1.5$ mm (black solid curve); $d = 5$ mm (blue solid curve). **c)** Dispersion of the edge modes when the response of the ferrite is regularized with a spatial-frequency cutoff $k_{max}$ (dashed curves) superimposed on the dispersion of the edge waveguide of panel a) for an air gap thickness $d = 1/k_{max}$ (solid curves). The color code is as in panel b).

The magnetically biased ferrite is characterized by a relative permittivity $\varepsilon_F = \varepsilon'_F (1 + i \tan \delta)$, with $\tan \delta$ being the electric loss tangent of the ferrite, and a frequency-dependent relative permeability tensor of the form $\bar{\bar{\mu}} = \begin{pmatrix} \mu_{11} & \mu_{12} & 0 \\ \mu_{21} & \mu_{22} & 0 \\ 0 & 0 & 1 \end{pmatrix}$, with

$$\mu_{11} = \mu_{22} = 1 + \frac{\omega_g (\omega_l - i\omega\alpha)}{(\omega_l - i\omega\alpha)^2 - \omega^2} \quad \text{and} \quad \mu_{12} = -\mu_{21} = -\frac{i\omega\omega_g}{(\omega_l - i\omega\alpha)^2 - \omega^2},$$

where $\alpha$ is a damping factor that determines the resonance linewidth [47]. Here, $\omega_l = \gamma B_z$ is the Larmor precession frequency that depends on the static magnetic bias $\mathbf{B}_0 = +B_z \hat{\mathbf{z}}$. The gyromagnetic ratio is given by $\gamma = e/m_e = 1.759 \times 10^{11}$ C/Kg, with *e* being the elementary charge and $m_e$ the electron mass. The Larmor frequency for the magnetization is given by $\omega_g = \mu_0 M_s \gamma$, where $M_s$ is the saturation magnetization. We consider a soft ferrite [48] with permittivity $\varepsilon'_F = 14.52$ and electric loss tangent $\tan \delta = 8 \times 10^{-5}$ that is magnetized on average by a static field $B_z = 0.16$ T. The corresponding Larmor frequencies are $\omega_l/(2\pi) = 4.43$ GHz and $\omega_g/(2\pi) = 5.18$ GHz. The damping factor is equal to $\alpha = 0.002$. The biasing field is created by a permanent bonded neodymium magnet [49].

The transverse electric (TE) polarized ferrite bulk modes, with $E_z \neq 0$, $H_z = 0$ and $\partial/\partial z = 0$, can be determined from the characteristic equation $\varepsilon_F \mu_{ef} (\omega/c)^2 = k^2$, where $c$ is the light speed in vacuum, $k^2 = \mathbf{k} \cdot \mathbf{k}$ with $\mathbf{k}$ being the wave vector, and $\mu_{ef} = \left(\mu_{11}^2 + \mu_{12}^2\right)/\mu_{11}$. The band diagram of the ferrite in the limit of vanishing loss is depicted in Fig. 1b. As seen, there is a band-gap defined by the frequency interval $\omega_L < \omega < \omega_H$, with $\omega_L = \omega_l \sqrt{\omega_g/\omega_l + 1}$ ($\omega_L = 2\pi \times 6.53$ GHz) and $\omega_H = \omega_l + \omega_g$ ($\omega_H = 2\pi \times 9.61$ GHz). On the other hand, the metal is taken as aluminum, which in the microwave range can be regarded as a perfect electric conductor. Thus, the metal shares the same band gap as the ferrite.

When two inequivalent topological materials share a common bandgap, their interface supports unidirectional edge states topologically protected against backscattering. The net number of edge modes is determined by the difference of the topological charges in the material [3, 22-24]. As already discussed in the introduction, systems with a continuous translation symmetry have an ill-defined topology. The topology can be regularized with a high-frequency cutoff $k_{max}$ that ensures that when $k \to \infty$ the material response becomes reciprocal [25, 50]. It should be noted that the suppression of the material response for $k \to \infty$ is not a mere mathematical trick but is rather a consequence of the Kramers-Kronig relations for spatially dispersive media and of the Riemman-Lebesgue lemma [51].

Using the topological band theory [25, 50, 52], it may be shown that the gap Chern number of the ferrite is $\mathcal{C}_{\text{gap,Fe}} = +1$ for a finite $k_{\max}$. The Chern number is robust to the presence of significant material loss [53, 54]. In contrast, without the cutoff the gap Chern number is $\mathcal{C}_{\text{gap,Fe}}\big|_{\text{local}} = \text{sgn}(\omega_g)\left(1+\left|\frac{\omega_g}{\omega_l}\right|\right)^{-1/2} \approx 0.68$ [25], which evidently is not an integer. Thus, the spatial cutoff is essential to regularize the topology of the material, which otherwise is ill-defined.

The regularization concept can be geometrically explained as follows. Figure 2a depicts a torus with vanishing inner radius. Any cross-section of the torus that contains the symmetry axis consists of two kissing circles. It is evident that the surface is not differentiable at the central point. Furthermore, the number of holes (genus) is ill-defined because the two kissing circles touch a single point. Thus, the considered surface has an ill-defined topology. This problem can be fixed with an infinitesimally small perturbation of the original object. For example, by increasing the inner radius of the object in Fig. 2a by an infinitesimal amount, the resulting surface becomes topologically equivalent to a torus (Fig. 2b) with genus $g=1$. Alternatively, the top and bottom caps can be separated by an infinitesimal distance near the singular point, yielding an object topologically equivalent to a sphere (Fig. 2c) with genus $g=0$. Thus, even infinitesimally small perturbations of the original system, can lead to a well-defined surface topology. Similarly, the topology of the ferrite becomes well-defined for an arbitrarily small value of $1/k_{\max}$. Curiously, the described geometrical ideas also highlight that the regularized topology may depend on the regularization procedure.

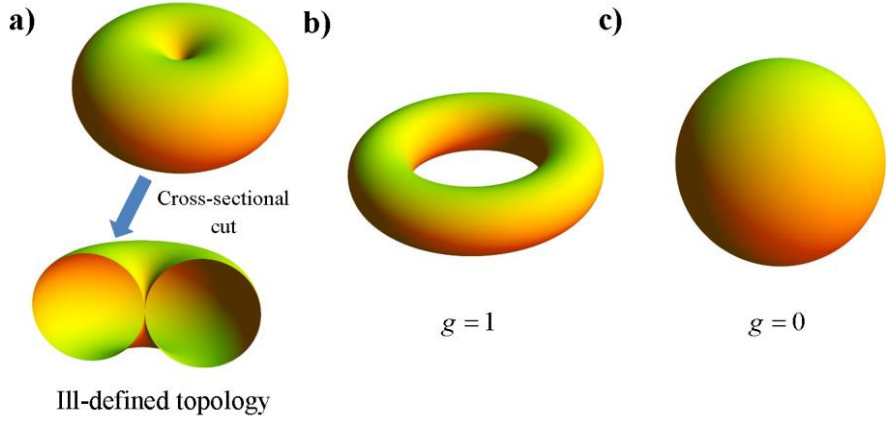

Fig. 2 Geometrical illustration of the concept of an ill-defined topology. **a)** A torus with a vanishingly small inner radius has a cross section consisting of two-kissing circles and thus it is topologically ill-defined. The topology can be regularized with an arbitrarily weak perturbation. For example, by increasing slightly the inner radius of the object one obtains a geometrical shape topologically equivalent to a torus ($g=1$) (panel **b**), whereas by separating the top and bottom caps by an infinitesimal distance at the singular point the object becomes topologically equivalent to a sphere ($g=0$) (panel **c**).

The gap Chern number of the metal vanishes due to reciprocity, and therefore the metal is topologically trivial. Hence, from the bulk-edge correspondence principle, it is expected that the metal-ferrite waveguide supports a single gapless unidirectional guided mode ($\delta\mathcal{C} = \left| \mathcal{C}_{\text{gap,Fe}} - \mathcal{C}_{\text{gap,Al}} \right| = 1$).

Figure 1c depicts the calculated edge mode dispersion for different values of $k_{\max}$ (dashed curves). The spatial dispersion arising due to the spatial cutoff is rigorously taken into account in the simulation using the same methods as in Ref. [23]. As seen, in agreement with the bulk-edge correspondence principle, the material interface supports exactly a gapless unidirectional edge-state, for any finite value of $k_{\max}$ (without an air gap in between the two materials).

The idealized $k_{\max}$-cutoff can be imitated with a "synthetic cutoff" that is implemented in practice by inserting a small air gap in between the two materials. The equivalent cutoff is related to the thickness $d$ as $k_{\max} \sim 1/d$ [23]. As seen, in Fig. 1c (solid lines, also repeated in Fig. 1b) the synthetic cutoff imitates fairly well the idealized $k_{max}$-cutoff, especially near the top edge of the bandgap. The dispersion equation of the edge modes with the synthetic cutoff is derived in Appendix A.

As seen in Fig. 1b, for any nonzero $d$, the ferrite-metal waveguide supports a single unidirectional edge state in most of the gap, consistent with the bulk edge correspondence. However, different from Fig. 1c (dashed lines), with the synthetic cutoff the edge mode is not strictly gapless. In fact, the dispersion of the edge state may not span the entire gap $\omega_L < \omega < \omega_H$ for $d \neq 0$, and when it does there is a second edge mode for frequencies near $\omega_L$ [see blue curve in Fig. 1b]. Hence, even though the synthetic cutoff regularization reproduces the main features of the idealized $k_{max}$-cutoff, the analogy is not strictly perfect.

For small values of $d$, the group velocity of the edge mode is negative ($v_{g,x} = \partial \omega_k / \partial k_x < 0$) in most of the bandgap, so that the energy flows along the $-x$-direction. Furthermore, for a small $d$ the edge mode in most of the gap is associated with a positive $k_x$ and thereby is typically a backward wave. Interestingly, as $d \to 0$ the group velocity approaches zero and the edge mode propagation is exactly suppressed when the gap is removed. This property may be attributed to the ill-defined topology of the ferrite without a spatial cutoff. Note that the same type of behavior occurs when $1/k_{max} \to 0$ (see Fig. 1c).

## III. Experimental Results

The results of the previous section show that the synthetic cutoff restores (at least partially) the bulk-edge correspondence and ensures that the waveguide supports a unidirectional edge mode. Here, we report the experimental verification of that property.

Our prototype consists of a hollow aluminum waveguide partially filled with a magnetically biased ferrite block as shown in Fig. 3a. The magnetic bias is created by a permanent magnet placed at the back of the waveguide (not visible in Fig. 3a). The ferrite block has dimensions $L_x = 150$ mm, $L_y = 50$ mm and $L_z = 10$ mm. The synthetic cutoff is implemented with an air gap with thickness $d = 5$ mm. In the first part of the experiment, the air gaps in between the lateral (left and right) walls of the metallic guide and the ferrite block are filled with an absorber (Eccosorb LS-26).

The edge mode is excited with small monopole antennas oriented along the $z$-direction in the middle of the air gap of the bottom lateral wall. The monopoles are fed by SMA cables, denoted in Fig. 3a by ports 1 and 2. The top and bottom walls ($z = const.$) are metallic plates that exactly emulate the two-dimensional scenario studied Sect. II, with $\mathbf{E} = E_z \hat{\mathbf{z}}$ and $\partial/\partial z = 0$.

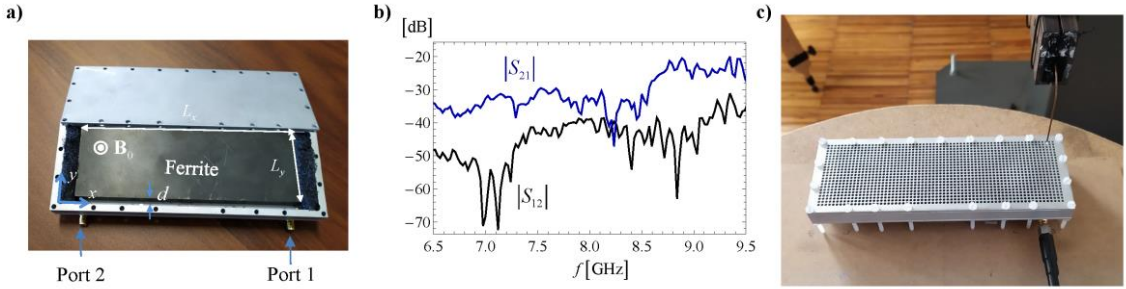

Fig. 3 **a)** Prototype of the unidirectional edge waveguide. The structure is formed by a hollow aluminum waveguide partially filled with a magnetically-biased ferrite block with dimensions $L_x = 150$ mm, $L_y = 50$ mm and $L_z = 10$ mm. The ferrite is biased by a permanent magnet placed at the back of the waveguide (not shown in the figure). The ferrite block is separated from the bottom lateral metallic wall by an air gap with thickness $d = 5$ mm. Absorbent blocks are placed at the left and right sides of the ferrite block. The structure is excited by small monopole antennas fed through ports 1 and 2. **b)** Measured amplitude of the scattering parameters $S_{12}$ and $S_{21}$. The transmission level from port 1 to port 2 is larger due to the excitation of an edge mode at the ferrite-air-metal interface. **c)** Edge mode measurement apparatus. The top cover of the waveguide is replaced by a metallic plate drilled with small holes. The holes have 2 mm diameter and are separated from each other by 1.5 mm in each direction. The electric field is measured at 1 mm from the surface of the hole grid using a small metallic probe that is moved using a robotic arm and is connected to a vector network analyzer (R&S ZVB20).

The theoretical dispersion of the edge mode is given by the blue curve in Fig. 1b for $d = 5$ mm. To verify the edge mode propagation, we measured the scattering parameters in the bandgap of the ferrite. The results are depicted in Fig. 3b and reveal a considerable asymmetry in the scattering parameters, with the transmission from port 1 to port 2 being significantly stronger than the transmission from port 2 to port 1. The reflection parameters $S_{11}$ and $S_{22}$ [not shown] are similar for both ports. Note that the transmission level is generally very low because of the large mismatch between the input impedance of the monopole antennas and the impedance of the feeding lines. As neither the ferrite nor the metal supports bulk states in the bandgap, the enhanced transmission level $S_{21}$ in the bandgap must be due to the excitation of the edge mode that flows along the –x-direction, which is the only possible radiation channel.

### A. Edge mode field profile

We measured the spatial distribution of the edge mode electric field using a near-field scanner. The near-field scanner is based on a robotic arm that controls the position of a small metallic probe oriented along the z-direction connected to a vector network analyzer (R&S ZVB20). In order to measure the electric field distribution of the edge mode, the top cover of the waveguide was replaced by a metallic plate drilled with an array of subwavelength circular holes. The holes have 2 mm diameter and are separated from each other by 1.5 mm in each direction. The subwavelength-sized holes allow for some leakage of the edge mode energy and thereby the external characterization of the fields, without interfering significantly with the propagation inside the waveguide. We considered two frequencies inside the bandgap, $f_1 = 7$ GHz and $f_2 = 7.25$ GHz. The measurement apparatus is shown in Fig. 3c. The electric field distributions for port 1 and port 2 excitations are shown in Figs. 4a-b and Figs. 4c-d, respectively.

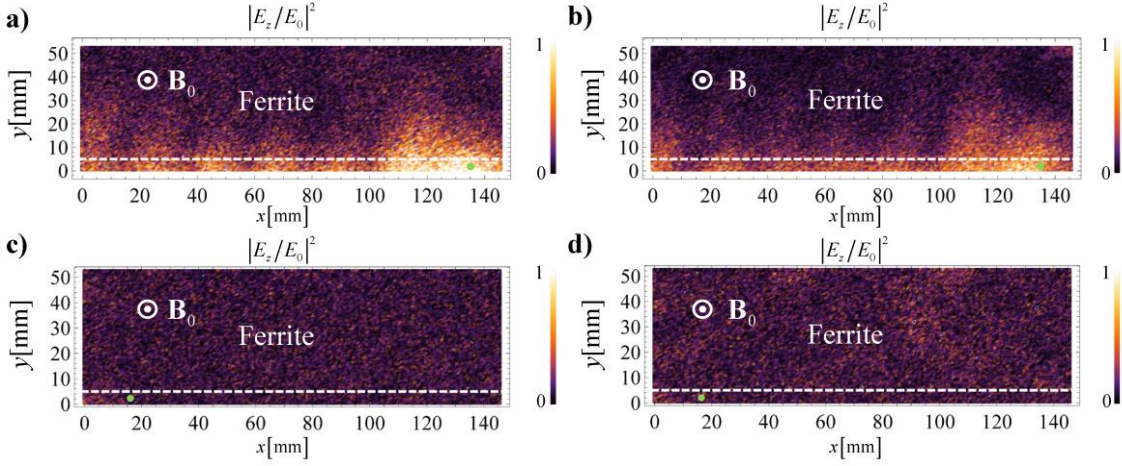

Fig. 4 Measured electric field distributions $|E_z/E_0|^2$ near the top metallic plate drilled with the hole array. **a)** and **b)** Port 1 excitation (propagation from the right to the left) for an oscillation frequency $f_1 = 7$ GHz and $f_2 = 7.25$ GHz, respectively. **c)** and **d)** are similar to **a)** and **b)** but for a port 2 excitation (propagation from the left to the right). In all panels, the small green dot represents the position of the excited monopole antenna.

Figures 4a-b show that when the waveguide is excited at port 1 the radiation flows along the air gap towards port 2 for both frequencies. In contrast, when the waveguide is excited at port 2 no field propagates towards port 1. Thus, the near-field scan confirms the nonreciprocal transmission in the waveguide, in agreement with the scattering matrix results of Fig. 3b. Moreover, since the measured electric field for a port 1 excitation is mostly concentrated in the ferrite-air-metal junction and propagates towards the –x-direction, it is evidently associated with the unidirectional edge mode [see blue curve of Fig. 1b]. The ripples in the field distribution along the propagation path can be attributed to two factors: *i)* the static magnetic field created by the magnets is not homogeneous in the ferrite block (the inhomogeneous biasing disturbs the permeability of the ferrite and thus modifies the propagation characteristics of the edge mode); *ii)* the field above the perforated metal plate (where the measurement is done) is far less confined than inside the waveguide. The latter point is illustrated in Fig. 5a, which depicts a snapshot of the measured electric field for a port 1 excitation at the frequency $f_2 = 7.25$ GHz. Indeed, above the top plate the edge mode spreads along the surface due to diffraction effects.

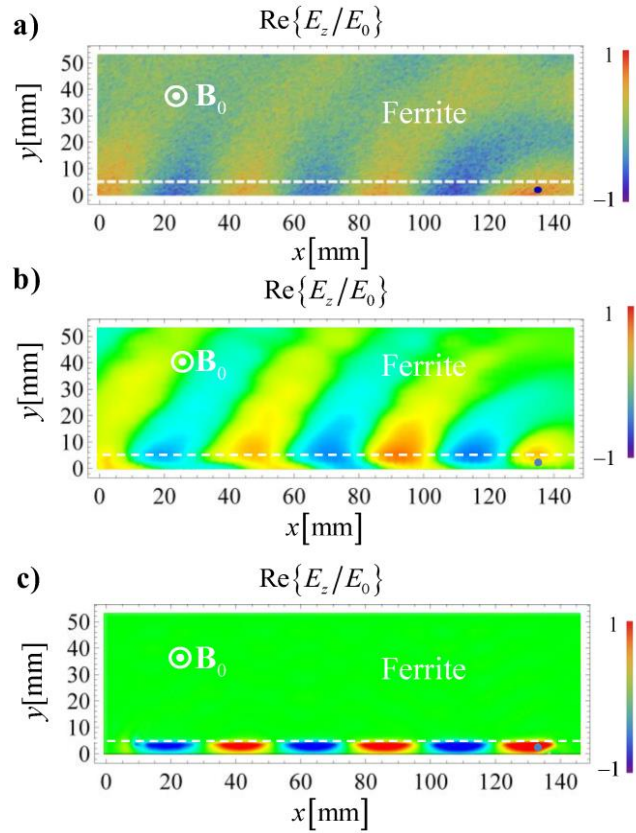

Fig. 5 **a)** Time snapshot of the measured normalized electric field distribution $\text{Re}\{E_z/E_0\}$ at the surface of the perforated metal cover of the waveguide for a port 1 excitation and for $f_2 = 7.25$ GHz. **b)** similar to **a)** but for the full-wave simulation [55]. **c)** similar to **b)** but the electric field is monitored inside the waveguide. In all panels, the small blue dot on the bottom right corner represents the position of the monopole antenna.

We also simulated the response of the ferrite-metal "edge waveguide" using the commercial software CST Studio Suite [55] (Fig. 5b-c). As seen in Fig. 5b, the simulation results follow closely the experimental results depicted in Fig. 5a. The simulation of Fig. 5c shows a time snapshot of the electric field inside the waveguide. In this case, the fields are strongly confined to the air gap near the bottom metallic wall. Furthermore, the full-wave results confirm that there is a unique unidirectional edge mode propagating in the waveguide, in agreement with the bulk-edge correspondence. The edge mode wavelength estimated from the experimental results is $\lambda_g \approx 49.5$ mm, whereas the value obtained from the full-wave simulations and from the modal equation (A2) is $\lambda_g \approx 55$ mm. The difference between the two values is attributed to the lack of homogeneity of the bias magnetic field in the experiment. For completeness, we note that in the scenario of Fig. 5 the edge mode is a forward wave: the phase and group velocities are both directed along –$x$.

### B. *Topological energy sink*

As previously discussed, the synthetic spatial cutoff determined by the air gap is essential to guarantee that the system supports a unidirectional edge mode. The air gap effectively regularizes the response of the ferrite and its topology, ensuring that the bulk edge correspondence is applicable in the spectral range of interest. On the other hand, when the air gap is removed the topology of the ferrite becomes ill-defined and the waveguide does not support edge modes. As shown next, this peculiar behavior creates the conditions for the formation of a topological singularity (energy sink) where the wave propagation is abruptly halted [21, 30-37].

To demonstrate the formation of an energy sink in the considered structure we removed the absorber in the left-hand side of the ferrite block. The air gap that now separates the ferrite from the metal at this lateral interface creates an edge waveguide with a well-defined topology supporting a unidirectional edge-mode. Thus, when the metallic waveguide is excited by the monopole antenna, the unidirectional guided mode can go around the sharp corner of the ferrite block without any backscattering and flow along the lateral left wall. Importantly, at the top lateral wall there is no air separation between the ferrite and the metal. Without the synthetic cutoff the ill-defined topology of the ferrite reveals itself. In fact, the absence of edge states at the top lateral wall and the existence of an edge state in the left lateral wall is a clear violation of the bulk-edge correspondence. The point where the ferrite block meets the top lateral wall corresponds to the transition between the well-defined topology and the ill-defined topology. The edge mode comes to an immediate halt at that point leading to the formation of an energy sink that dissipates all the incoming energy.

In order to experimentally verify these ideas, the structure was excited at port 2 for two frequencies inside the common bandgap of the materials, $f_1 = 7\text{ GHz}$ and $f_2 = 7.25\text{ GHz}$. The electric field was measured at the surface of the perforated metal plate.

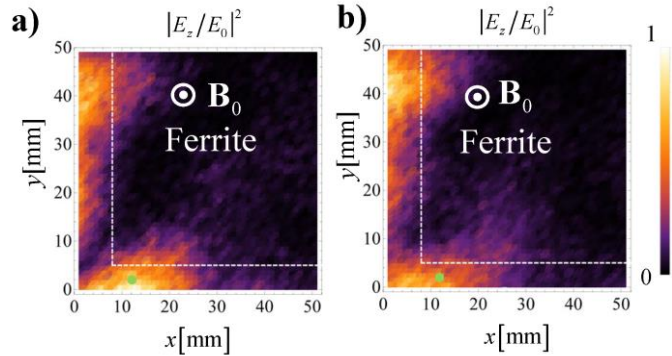

Fig. 6 Normalized electric field distribution $\left|E_z/E_0\right|^2$ at the surface of the drilled metal plate for a port 2 excitation at the oscillation frequency **a)** $f_1 = 7\text{ GHz}$ and **b)** $f_2 = 7.25\text{ GHz}$. The small green dot represents the position of the monopole antenna.

The experimental results depicted in Fig. 6 demonstrate that the excited edge mode propagates without back-reflections around the sharp corner of the ferrite block, coming to a halt at the top corner of the structure. This can be seen in Fig. 6, which shows a large field concentration near the top metallic wall of the waveguide. All the energy coupled to the edge mode is ultimately absorbed at the energy sink, leading to a massive field enhancement [21]. Note that the field intensity near the sink is much stronger than in any other point in the lateral channel, being comparable to the field measured near the monopole antenna. In principle, the contrast between the field at the hotspot and the field in the lateral channel is even more pronounced inside the waveguide.

## IV. Conclusions

Nonreciprocal photonic systems with a continuous translation symmetry typically have an ill-defined topology. The scattering of unidirectional modes in such systems is not constrained by the bulk edge correspondence. In this work, we experimentally verified that the topology of a ferrite-metal waveguide can be effectively regularized with a synthetic cutoff determined by a small air gap. Furthermore, it was experimentally demonstrated that by pairing two waveguides, one with a well-defined topology and another with an ill-defined topology, it is possible to bring an edge mode to an immediate halt at a topological singularity where the electromagnetic fields are massively enhanced. Our findings suggest that photonic systems with ill-defined topologies are unique platforms for extreme wave phenomena.

**Acknowledgements**

This work was partially funded by the Institution of Engineering and Technology (IET) under the A F Harvey Research Prize 2018, by the Simons Foundation, by Fundação para Ciência e a Tecnologia under Project PTDC/EEITEL/4543/2014 and by Instituto de Telecomunicações under Project No. UID/EEA/50008/2020. D. E. Fernandes acknowledges support by FCT, POCH, and the cofinancing of Fundo Social Europeu under the fellowship SFRH/BPD/116525/2016. R. A. M. Pereira acknowledges FCT for the Ph.D. grant SFRH/BD/145024/2019. T. A. Morgado acknowledges FCT for research financial support with reference CEECIND/04530/2017 under the CEEC Individual 2017, and IT-Coimbra for the contract as an assistant researcher with reference CT/No. 004/2019-F00069. SL acknowledges FCT and IT-Coimbra for the research financial support with reference DL 57/2016/CP1353/CT000.

## Appendix A: The dispersion equation of the edge modes

Here we derive the modal equation of the edge modes that propagate in the waveguide depicted in Fig. 1a. To do so, we write the electric field of the TE-waves as a superposition of decaying exponentials along the *y*-direction, so that

$$E_z(x,y,\omega) = E_0 e^{ik_x x} \begin{cases} e^{-\gamma_F(y-d)}, & y > d \\ \dfrac{\sinh(\gamma_0 y)}{\sinh(\gamma_0 d)}, & 0 < y < d \end{cases}, \quad (A1)$$

being $k_x$ the propagation constant of the edge mode (along *x*). Note that $E_z$ is continuous and vanishes at the metal wall ($y=0$). Here, the attenuation constants in the ferrite and air are $\gamma_F = \sqrt{k_x^2 - \mu_{ef}\varepsilon_F(\omega/c)^2}$ and $\gamma_0 = \sqrt{k_x^2 - (\omega/c)^2}$.

The magnetic field is obtained from the Maxwell equations using $\mathbf{H} = \dfrac{1}{i\omega\mu_0} \bar{\bar{\mu}}^{-1} \cdot \nabla \times \mathbf{E}$.

Enforcing the continuity of $H_x = \dfrac{1}{i\omega\mu_0} \dfrac{1}{\mu_{11}^2 + \mu_{12}^2} \left( \mu_{11}\partial_y E_z + \mu_{12}\partial_x E_z \right)$ at $y = d$, we obtain the desired dispersion equation:

$$\gamma_F + \mu_{ef}\gamma_0 \coth(\gamma_0 d) - \dfrac{\mu_{12}}{\mu_{11}} ik_x = 0. \quad (A2)$$

## References

[1] F. D. M. Haldane and S. Raghu, "Possible Realization of Directional Optical Waveguides in Photonic Crystals with Broken Time-Reversal Symmetry", *Phys. Rev. Lett.* **100**, 013904 (2008).
[2] S. Raghu and F. D.M. Haldane, "Analogs of quantum-Hall-effect edge states in photonic crystals", *Phys. Rev. A* **78**, 033834 (2008).
[3] L. Lu, J. D. Joannopoulos, and M. Soljačić, "Topological photonics", *Nat. Photonics* **8**, 821 (2014).

[4] L. Lu, J. D. Joannopoulos, and M. Soljačić, "Topological states in photonic systems", *Nat. Phys.* **12**, 626 (2016).

[5] L. E. Zhukov and M. E. Raikh, "Chiral electromagnetic waves at the boundary of optical isomers: Quantum Cotton-Mouton effect", *Phys. Rev. B* **61**, 12842 (2000).

[6] Z. Yu, G. Veronis, Z. Wang, and S. Fan, "One-Way Electromagnetic Waveguide Formed at the Interface Between a Plasmonic Metal Under a Static Magnetic Field and a Photonic Crystal", *Phys. Rev. Lett.* **100**, 023902 (2008).

[7] Z. Wang, Y. Chong, J. D. Joannopoulos, and M. Soljačić, "Observation of unidirectional backscattering immune topological electromagnetic states", *Nature* **461**, 772 (2009).

[8] M. C. Rechtsman, J. M. Zeuner, Y. Plotnik, Y. Lumer, D. Podolsky, F. Dreisow, S. Nolte, M. Segev, and A. Szameit, "Photonic Floquet topological insulators", *Nature* **496**, 196 (2013).

[9] W. Gao, M. Lawrence, B. Yang, F. Liu, F. Fang, B. Béri, J. Li, and S. Zhang, "Topological Photonic Phase in Chiral Hyperbolic Metamaterials", *Phys. Rev. Lett.* **114**, 037402 (2015).

[10] T. Ma, A. B. Khanikaev, S. H. Mousavi, and G. Shvets, "Guiding Electromagnetic Waves around Sharp Corners: Topologically Protected Photonic Transport in Metawaveguides", *Phys. Rev. Lett.* **114**, 127401 (2015).

[11] M. Hafezi, E. A. Demler, M. D. Lukin, and J.M. Taylor, "Robust optical delay lines with topological protection", *Nat. Phys.* **7**, 907 (2011).

[12] O. Luukkonen, U. K. Chettiar and N. Engheta, "One-way waveguides connected to one-way loads", *IEEE Antennas and Wireless Propag. Lett.* **11**, 1398 (2012).

[13] A. B. Khanikaev, S. H. Mousavi, W.-K. Tse, M. Kargarian, A. H. MacDonald, and G. Shvets, "Photonic topological insulators", *Nat. Mater.* **12**, 233 (2013).

[14] A. Slobozhanyuk, S. H. Mousavi, X. Ni, D. Smirnova, Y. S. Kivshar, and A. B. Khanikaev, "Three-dimensional all-dielectric photonic topological insulator", *Nat. Photonics* **11**, 130 (2017).

[15] C. He, X.-C. Sun, X.-P. Liu, M.-H. Lua, Y. Chen, L. Feng, and Y.-F. Chen, "Photonic topological insulator with broken time-reversal symmetry", *Proc. Natl. Acad. Sci. U.S.A.* **113**, 4924 (2016).

[16] W.-J. Chen, Z.-Q. Zhang, J.-W. Dong, and C.T. Chan, "Symmetry-protected transport in a pseudospin-polarized waveguide", *Nat. Commun.* **6**, 8183 (2015).

[17] R. Fleury, A. Khanikaev, and A. Alù, "Floquet topological insulators for sound", *Nat. Commun.* **7**, 11744 (2016).

[18] M. G. Silveirinha, "$Z_2$ Topological Index for Continuous Photonic Materials", *Phys. Rev. B* **93**, 075110 (2016).

[19] M. G. Silveirinha, "PTD symmetry protected scattering anomaly in optics", *Phys. Rev. B* **95**, 035153 (2017).

[20] M. G. Silveirinha, "Topological Angular Momentum and Radiative Heat Transport in Closed Orbits", *Phys. Rev. B* **95**, 115103 (2017).

[21] D. E. Fernandes, M. G. Silveirinha, "Topological origin of electromagnetic energy sinks", *Phys. Rev. Appl.* **12**, 014021 (2019).

[22] Y. Hatsugai, "Chern number and edge states in the integer quantum Hall effect", *Phys. Rev. Lett.* **71**, 3697, (1993).

[23] M. G. Silveirinha, "Bulk-edge correspondence for topological photonic continua", *Phys. Rev. B* **94**, 205105 (2016).

[24] M. G. Silveirinha, "Proof of the bulk-edge correspondence through a link between topological photonics and fluctuation-electrodynamics", *Phys. Rev. X*, **9**, 011037 (2019).

[25] M. G. Silveirinha, "Chern Invariants for Continuous Media", *Phys. Rev. B* **92**, 125153 (2015).

[26] S. Buddhiraju, Y. Shi, A. Song, C. Wojcik, M. Minkov, I. A. D. Williamson, A. Dutt, and S. Fan, "Absence of unidirectionally propagating surface plasmon-polaritons in nonreciprocal plasmonics", *Nat. Commun.* **11**, 674 (2020).

[27] S. A. H. Gangaraj and F. Monticone, "Do Truly Unidirectional Surface Plasmon-Polaritons Exist?", *Optica* **6**, 9 (2019).

[28] S. Pakniyat, G.W. Hanson and S. A. H. Gangaraj, "Chern Invariants of Topological Continua; a Self-Consistent Nonlocal Hydrodynamic Model arXiv:2110.10768v1.

[29] S. A. H. Gangaraj and F. Monticone, "Physical Violations of the Bulk-Edge Correspondence in Topological Electromagnetics", *Phys. Rev. Lett.* **124**, 153901 (2020).

[30] B. Lax, and K. J. Button, "New Ferrite Mode Configurations and Their Applications", *J. Appl. Phys.* **26**, 1186 (1955).

[31] M. Kales, "Topics in guided-wave propagation in magnetized ferrites", *Proc. IRE* **44**, 1403 (1956).

[32] A. Bresler, "On the $TE_{n0}$ modes of a ferrite slab loaded rectangular waveguide and the associated thermodynamic paradox", *IRE Trans. Microwave Theory Tech.* **8**, 81 (1960).

[33] A. Ishimaru, "Unidirectional waves in anisotropic media and the resolution of the thermodynamic paradox", *Tech. Rep.* (Washington Univ. Seattle, 1962).

[34] G. Barzilai and G. Gerosa, "Rectangular waveguides loaded with magnetised ferrite, and the so-called thermodynamic paradox", *Proc. IEE* **113**, 285 (1966).

[35] U. K. Chettiar, A. R. Davoyan, and N. Engheta, "Hotspots from nonreciprocal surface waves", *Opt. Lett.* **39**, 1760 (2014).

[36] L. Shen, Z. Wang, X. Deng, J.-J. Wu, and T.-J. Yang, "Complete trapping of electromagnetic radiation using surface magnetoplasmons", *Opt. Lett.*, **40**, 1853, (2015).

[37] K. L. Tsakmakidis, L. Shen, S. A. Schulz, X. Zheng, J. Upham, X. Deng, H. Altug, A. F. Vakakis, and R. W. Boyd, "Breaking Lorentz reciprocity to overcome the time-bandwidth limit in physics and engineering", *Science* **356**, 1260 (2017).

[38] F. J. Garcia-Vidal, and J. B. Pendry, "Collective Theory for Surface Enhanced Raman Scattering", *Phys. Rev. Lett.* **77**, 1163 (1996).

[39] K. Kneipp, Y. Wang, H. Kneipp, L. Perelman, I. Itzkan, R. Dasari, and M. Feld, "Single Molecule Detection Using Surface-Enhanced Raman Scattering (SERS)", *Phys. Rev. Lett.* **78**, 1667 (1997).

[40] A. Campion and P. Kambhampati, "Surface-enhanced Raman scattering", *Chem. Soc. Rev.* **27**, 241 (1998).

[41] C. K. Chen, T. F. Heinz, D. Ricard, and Y.R. Shen, "Surface-enhanced second-harmonic generation and Raman scattering", *Phys. Rev. B* **27**, 1965 (1983).

[42] A. Bouhelier, M. Beversluis, A. Hartschuh, and L. Novotny, "Near-Field Second-Harmonic Generation Induced by Local Field Enhancement", *Phys. Rev. Lett.* **90**, 013903 (2003).

[43] U. K. Chettiar and N. Engheta, "Optical frequency mixing through nanoantenna enhanced difference frequency generation: Metatronic mixer", *Phys. Rev. B* **86**, 075405 (2012).

[44] S. Kim, J. Jin, Y.-J. Kim, I.-Y. Park, Y. Kim, and S.-W. Kim, "High-harmonic generation by resonant plasmon field enhancement". *Nature* **453**, 757 (2008).

[45] J. A. Schuller, E. S. Barnard, W. Cai, Y. C. Jun, J. S. White, and M. L. Brongersma, "Plasmonics for extreme light concentration and manipulation", *Nature Mater.* **9**, 193 (2010).

[46] H. A. Atwater, and A. Polman "Plasmonics for improved photovoltaic devices", *Nature Mater.* **9**, 205 (2010).

[47] D. M. Pozar, *Microwave Engineering,* 3rd Edition, Hoboken, NJ :Wiley, 2005

[48] https://www.skyworksinc.com/Products/Technical-Ceramics/TTVG-1850

[49] https://www.calamit.es/imanes-permanentes/bounded-de-neodimio/imanes-inyectados-comprimidos-neodimio.php

[50] M. G. Silveirinha, "Topological classification of Chern-type insulators by means of the photonic Green function", *Phys. Rev. B* **97**, 115146 (2018).

[51] V. M. Agranovich and V. Ginzburg, *Crystal Optics with Spatial Dispersion, and Excitons,* Springer Series in Solid-State Sciences, 1984.

[52] S. A. H. Gangaraj, and G.W. Hanson, "Topologically Protected Unidirectional Surface States in Biased Ferrites: Duality and Application to Directional Couplers", *IEEE Antennas and Wireless Propag. Lett.* **16**, 449 (2017).

[53] F. R. Prudêncio and M. G. Silveirinha, "First Principles Calculation of Topological Invariants of non-Hermitian Photonic Crystals", *Comm. Phys.* **3**, 221 (2020).

[54] M. G. Silveirinha, "Topological theory of non-Hermitian photonic systems", *Phys. Rev. B* **99**, 125155 (2019).

[55] SIMULIA - CST Studio Suite https://www.3ds.com/products-services/simulia/products/cst-studio-suite